\documentclass[preprint,prd,nofootinbib,tightenlines,amsmath,amssymb]{revtex4}
\usepackage{enumerate}
\usepackage{multirow}
\usepackage[utf8]{inputenc}
\usepackage{anysize}
\usepackage{textcomp}
\usepackage{epsfig}
\usepackage{graphics,color} 
\usepackage{verbatim}
\usepackage{afterpage}
\usepackage{amsmath}    
\usepackage{soul}
\usepackage{xcolor,cancel}
\DeclareUnicodeCharacter{00A0}{ }

\usepackage{blindtext}

\catcode`\@=11
\def\sla@#1#2#3#4#5{{%
 \setbox\z@\hbox{$\m@th#4#5$}%
 \setbox\tw@\hbox{$\m@th#4#1$}%
 \dimen4\wd\ifdim\wd\z@<\wd\tw@\tw@\else\z@\fi
 \dimen@\ht\tw@
 \advance\dimen@-\dp\tw@ \advance\dimen@-\ht\z@
 \advance\dimen@\dp\z@
 \divide\dimen@\tw@ \advance\dimen@-#3\ht\tw@
 \advance\dimen@-#3\dp\tw@ \dimen@ii#2\wd\z@
 \raise-\dimen@\hbox to\dimen4{%
 \hss\kern\dimen@ii\box\tw@\kern-\dimen@ii\hss}%
 \llap{\hbox to\dimen4{\hss\box\z@\hss}}}}

\def\cpto{\mathrel {\vcenter {\baselineskip 0pt \kern 0pt
    \hbox{$H_{r.f.}$} \kern 0pt \hbox{$\longrightarrow$} }}}
\def\slashed#1{%
 \expandafter\ifx\csname sla@\string#1\endcsname\relax
{\mathpalette{\sla@/00}{#1}}
\fi}
\def\declareslashed#1#2#3#4#5{%
 \expandafter\def\csname sla@\string#5\endcsname{%
#1{\mathpalette{\sla@{#2}{#3}{#4}}{#5}}}}
 \catcode`\@=12
\declareslashed{}{/}{.08}{0}{D}
 \declareslashed{}{/}{.1}{0}{A}
 \declareslashed{}{/}{0}{-.05}{k}
 \declareslashed{}{/}{.1}{0}{\partial}
 \declareslashed{}{\not}{-.6}{0}{f}

\def\lsim{\mathrel {\vcenter {\baselineskip 0pt \kern 0pt
    \hbox{$<$} \kern 0pt \hbox{$\sim$} }}}
\def\gsim{\mathrel {\vcenter {\baselineskip 0pt \kern 0pt
    \hbox{$>$} \kern 0pt \hbox{$\sim$} }}}

\newcommand{\bea}{\begin{eqnarray}}
\newcommand{\eea}{\end{eqnarray}}

\begin{document}

\baselineskip=15pt
\preprint{}

\title{Consequences of R-Parity violating interactions for anomalies in $\bar B\to D^{(*)} \tau \bar \nu$ and $b\to s \mu^+\mu^-$}

\author{N.G. Deshpande$^1$\footnote{Electronic address: desh@uoregon.edu}, Xiao-Gang He$^{2,3,4}$\footnote{Electronic address: hexg@phys.ntu.edu.tw}}
\affiliation{
$^{1}$Institute of Theoretical Science, University of Oregon, OR 97403, USA\\
$^{2}$INPAC, Department of Physics and Astronomy, Shanghai Jiao Tong University, Shanghai 200240, China.\\
$^{3}$CTS, CASTS and Department of Physics, National Taiwan University, Taipei 10617. \\
$^{4}$National Center for Theoretical Sciences, Hsinchu 300, Taiwan
}

\date{\today}

\vskip 1cm
\begin{abstract}
We investigate the possibility of explaining the enhancement in semileptonic decays of $\bar B \to D^{(*)} \tau \bar \nu$, the anomalies induced by $b\to s\mu^+\mu^-$
in $\bar B\to (K, K^*, \phi)\mu^+\mu^-$ and violation of  lepton universality in $R_K = Br(\bar B\to K \mu^+\mu^-)/Br(\bar B\to K e^+e^-)$ within the framework of R-parity violating (RPV) MSSM. Exchange of down type right-handed squark coupled to quarks and leptons yield  interactions which are similar to leptoquark induced interactions that have been proposed to explain the $\bar B \to D^{(*)} \tau \bar \nu$ by tree level interactions and $b\to s \mu^+\mu^-$  anomalies by loop induced interactions, simultaneously. However, the Yukawa couplings in such theories have severe constraints from other rare processes in $B$ and $D$ decays. Although this interaction can provide a viable solution to $R(D^{(*)})$ anomaly, we show that with the severe constraint from 
$\bar B \to K \nu \bar \nu$, it is impossible to solve the anomalies in $b\to s \mu^+\mu^-$ process simultaneously. 
\end{abstract}

\pacs{PACS numbers: }

\maketitle

\section{Introduction}

Recent experimental data have shown deviations from standard model (SM) predictions in the ratio of $R(D^{(*)}) = Br(\bar B \to D^{(*)} \tau \nu)/Br(\bar B \to D^{(*)} l \nu)$ with $l = e,\;\mu$ and also in $b\to s \mu^+ \mu^-$ induced $B$ decays. Experimental values for $R(D^{(*)})$\cite{RD-1,RD-2,RD-3} are larger than the SM predictions\cite{RD-sm}. This anomalous effect is significant, at about 4$\sigma$ level\cite{HFAG}.
The anomalies due to $b\to s \mu^+\mu^-$ induced processes show up in\cite{b2s-1} $B \to (K,K^*,\phi)\mu^+\mu^-$ decays. The observed branching ratios in these decays are lower than SM predictions\cite{b2s-sm, b2s-the-1}. Also a deficit is shown in the ratio $R_K = Br(B\to K\mu^+\mu^-)/Br(B\to K e^+ e^-)$\cite{RK}. The SM predicts $R_K$ to be close to one, but experimental data give\cite{RK} $0.745^{+0.090}_{-0.074}\pm 0.036$. These effects are at 2 to 3 $\sigma$. 
Needless to say that these anomalies need to be further confirmed experimentally and we also need to understand SM predictions better. The latter processes involved are rare processes and therefore are sensitive to new physics. These anomalies have attracted a lot of theoretical attentions trying to solve the problems using new physics beyond SM\cite{RD-sm, b2s-sm, b2s-the-1, RD-the-1, desh-menon,  leptoquark, b2s-the-2, RD-b2s, bauer-neubert}.  
In this work we study the possibility of using R-parity violating interaction to solve these anomalies. Previously, R-parity violation was invoked to explain\cite{desh-menon} only $R(D^{(*)})$.
Exchange of down type right-handed squark coupled to quarks and leptons yield interactions, which are similar to leptoquark induced interactions that have been proposed to explain the $\bar B \to D^{(*)}\to \tau \bar \nu$ and $b\to s \mu^+\mu^-$ induced anomalies simultaneously\cite{bauer-neubert}. However, the Yukawa couplings have severe constraints from other rare processes in $B$ and $D$ decays. This interaction can provide a viable solution to $R^{(*)}$ anomaly. But with severe constraint from 
$\bar B \to K \nu \bar \nu$, it proves to be impossible to solve the anomalies induced by $b\to s \mu^+\mu^-$process.

The most general renomalizable R-parity violating terms in the superpotentials are\cite{R-parity}
\begin{eqnarray}
W_{RPV} = \mu_i L_i H_u + {1\over 2} \lambda_{ijk}L_i L_j E^c_k + \lambda^\prime_{ijk}L_i Q_j D^c_k + {1\over 2} \lambda{''}_{ijk}U^c_iD^c_jD^c_k\;,.
\end{eqnarray}

We will assume that $\lambda''$ term is zero to ensure proton stability. 
Since the processes we discuss involve leptons and quarks, the $\lambda'$ term should remain. In fact  the interactions induced by this term at the tree and one loop level can contribute to $\bar B \to D^{(*)} \tau \bar \nu$ and $b\to s \mu^+\mu^-$ induced processes. It is tempting to see if these interactions can solve the related anomalies already. Although a combination of $\lambda'$ and $\lambda$ terms can also contribute, the resulting operators are disfavored by $\bar B\to D^{(*)} \tau \bar \nu$ process. 

We shall limit ourselves to exchange of right-handed down type squark, $\tilde d^k_R$, which are expected to have the necessary ingredients to explain the anomalies in $B$ decays. This model is similar to the leptoquark exchange discussed by many authors\cite{b2s-the-2}, except a general leptoquark also has a right-handed couplings to $SU(2)_L$ singlets, which is forbidden in SUSY. 
These additional right-handed couplings turn out to be important for explaining the $g-2$ anomaly of muon, but do not play an essential  role in explaining the B anomalies discussing here.
The object of our paper is a careful consideration of the constraints from various $B$ and $D$ decays and analysie structure of Yukawa couplings $\lambda^\prime_{ijk}$ to see if the B anomalies can be resolved simultaneously. 
The paper by Bauer and Neubert\cite{bauer-neubert} is closest in spirit to our paper, but we are able to bring out the tension between different experimental constraints, and find that it is impossible to solve the $R(D^{(*)})$ and $b \to s \mu^+\mu^-$ anomalies simultaneously.

The $R(D^{(*)})$ and $b\to s\mu^+\mu^-$ anomalies occur at tree level and loop level in the SM, respectively. To simultaneously solve these anomalies using a simple set of beyond SM interactions faces more constraints\cite{RD-b2s, bauer-neubert} than just solving one of them as has been done in most of the studies. We find that by exchanging right-handed down type of squark, it is possible to solve the $R(D^{(*)})$ anomaly with tree interaction provided $\lambda^\prime_{33k}$ is sizable, of order $\sim 3$. For anomalies induced by $b\to s \mu^+\mu^-$, to obtain the right chirality for operators $O_9$, one needs to go to one loop level. The allowed
couplings $\lambda^\prime_{ijk}$ are constrained from various experimental data, such as $K \to \pi \nu \bar \nu$,
$\bar B \to K(K^*) \nu \bar \nu$ and $D^0 \to \mu^+\mu^-$.  The strongest constraint comes from $\bar B \to K(K^*) \nu \bar \nu$ making the model impossible to explain anomalies induced by $b\to s \mu^+\mu^-$.

\section{R-parity violating interactions and $\bar B\to D^{(*)} \tau \bar \nu$}

Expanding the $\lambda'$ term in terms of fermions and sfermions, we have
\begin{eqnarray}
L = \lambda^{\prime}_{ijk}\left [ \tilde \nu^i_L \bar d^k_R d^j_L + \tilde d^j_L \bar d^k_R\nu^i_L + \tilde d^{k*}_R \bar \nu^{ci}_L d^j_L- \tilde l^i_L \bar d^k_R u^j_L - \tilde u^j_L \bar d^k_R l^i_L - \tilde d^{k*}_R \bar l^{ci}_L u^j_L\right ]\;,
\end{eqnarray}
where the ``tilde'' indicates the sparticles, and ``c'' indicates charge conjugated fields.

Working in the basis where down quarks are in their mass eigenstates, $Q^T = (V^{KM\dagger}u_L, d_l)$, one replaces $u^j_L$ in the above 
by $(V^{KM\dagger}u_L)^j$. Here $V^{KM}$ is the Kobayashi-Maskawa (KM) mixing matrix for quarks. If experimentally, the mass eigenstate of neutrino are not identified, one does not need to insert the PMNS mixing matrix for lepton sector. The neutrinos in the above equation are thus in the weak eigenstates. For leptoquark interactions discussed in eq. (6) in Ref.\cite{bauer-neubert}, the reference seems to indicate that new parameters are involved due to rotation matrix $U_e$ in the lepton sector. However, since neutrinos are not in the mass basis in our work, it seems that provided we are always in the weak basis, no matrix is required in the lepton sector. We will assume sfermions are in their mass eigenstate basis. For a discussion of the choice of basis see Ref.\cite{R-parity}

Exchanging sparticles, one obtains the following four fermion operators at the tree level
\begin{eqnarray}
{\cal L}_{eff} &=& {\lambda'_{ijk}\lambda^{'*}_{i'j'k}\over 2 m^2_{\tilde
d^k_R}} \left [ \bar \nu^{i'}_L \gamma^\mu \nu^i_L \bar d^{j'}_L
\gamma_\mu d_L^j + \bar e^{i'}_L \gamma^\mu e^i_L 
(\bar u_L V^{KM})^{j'}\gamma_\mu (V^{KM\dagger}u_L)^j \right .\nonumber\\
&&\hspace{1.7cm} -\nu^{i'}_L \gamma^\mu e^i_L \bar d^{j'}_L \gamma_\mu (V^{KM\dagger}u_L)^j 
-\bar e^{i'}_L \gamma^\mu \nu^i_L (\bar u_LV^{KM})^{j'} \gamma_\mu
d^j_L\left .  \right ]\nonumber\\
&-&{\lambda'_{ijk}\lambda^{'*}_{i'jk'}\over 2m^2_{\tilde d^j_L} }
\bar \nu^{i'}_L \gamma^\mu \nu^i_L \bar d^k_R \gamma_\mu d^{k'}_R
-{\lambda'_{ijk}\lambda^{'*}_{i'jk'}\over 2m^2_{\tilde u^j_L}}
\bar e^{i'}_L \gamma^\mu e^i_L \bar d^k_R \gamma_\mu
d^{k'}_R \\
&-&{\lambda'_{ijk}\lambda^{'*}_{ij'k'}\over 2m^2_{\tilde e^i_L}}
 (\bar u_{L\beta} V^{KM})^{j'} \gamma^\mu (V^{KM\dagger}u_{L\alpha})^j \bar d^k_{R\alpha}
\gamma_\mu d^{k'}_{R\beta} - {\lambda'_{ijk}\lambda^{'*}_{ij'k'}\over 2m^2_{\tilde \nu^i_L}}
\bar d^{j'}_{L\beta} \gamma^\mu d^j_{L\alpha} \bar d^k_{R\alpha}
\gamma_\mu d^{k'}_{R\beta}, \nonumber \label{intr}
\end{eqnarray}
In the above $\alpha$ and $\beta$ are color indices.

At the tree level, besides the SM contributions to $\bar B\to D^{(*)} l \bar \nu$, there are also R-parity violating contributions, they are given by the term proportional to $ - (\lambda'_{l3k}\lambda^{\prime*}_{l'mk}/2 m^2_{\tilde
d^k_R})\bar l_L \gamma^\mu \nu^{l'}_L (\bar u_LV^{KM})^{m} \gamma_\mu b_L$ in the above equation.
Including the SM contributions one obtains\cite{desh-menon}
\begin{eqnarray}
H_{eff} &=& -{4G_F\over \sqrt{2}}V_{m3} (\delta^{l'}_l + \Delta^{l'm}_l) \bar l \gamma^\mu P_L \nu_{l'} \bar u^m\gamma_\mu P_L b_L\;,\nonumber\\
\Delta^{l,m}_{l'} &=& {\sqrt{2}\over 4 G_F} {\lambda^\prime_{l3k} \lambda^{\prime *}_{l' j' k}\over 2 m^2_{\tilde d^k_R}}{V_{mj'}\over V_{m3}} \;.\label{Heffbc}
\end{eqnarray}
where $V_{ij}$ are elements in $V^{KM}$.

Identifying different charged leptons in the final states, we find the ratio $R^{SM}_l(c) = Br(\bar B\to D^{(*)} l \nu)/Br(\bar B \to D^{(*)} l \nu)_{SM}$ of branching ratios compared with SM predictions to be given by
\begin{eqnarray}
&&R^{SM}_\tau(c) =  \vert \Delta_1^{3,2}\vert^2 + \vert \Delta_2^{3,2}\vert ^2 + \vert 1+\Delta_3^{3,2}\vert^2 \;,\nonumber\\
&&R^{SM}_\mu(c) = \vert \Delta_1^{2,2}\vert ^2 + \vert 1+\Delta_2^{2,2}\vert ^2  + \vert \Delta_3^{2,2}\vert ^2\;,\nonumber\\
&&R^{SM}_e(c) =  \vert 1+\Delta_1^{1,2}\vert ^2+\vert \Delta_2^{1,2}\vert ^2 +\vert \Delta_3^{1,2}\vert ^2 \;. \label{RSM}
\end{eqnarray}

One can define a similar quantity $R^{SM}_l(u)$ for $Br(\bar B\to (\rho, \pi) l \nu)/Br(\bar B \to (\rho, \pi) l \nu)_{SM}$ and
$Br(\bar B\to l \nu)/Br(\bar B \to l \nu)_{SM}$, and have 
\begin{eqnarray}
R^{SM}_l(u)={Br(\bar B\to (\rho, \pi) l \nu)\over Br(\bar B \to (\rho, \pi) l \nu)_{SM}}={Br(\bar B\to l \nu)\over Br(\bar B \to l \nu)_{SM}}\;.
\end{eqnarray}

Experimentally, $R^{SM}_{e}$ deviations from SM predictions is small, that is $R^{SM}_{e} \approx 1$, therefore we require $\Delta_i^{1,2}$, to be close to zero, which can be achieved by setting $\lambda^\prime_{1jk} = 0$,  so that no linear terms in $\Delta^{i,j}_k$ contribute to $\bar B\to D^{(*)} e \bar \nu_e$. 
No large deviation has been observed in $R^{SM}_\mu$. However in $b\to s\mu^+\mu^-$ induced anomalies involves $\mu$ couplings, we will bare in mind that effect may have some impact for $R^{SM}_\mu(c)$. One may even contemplate that 
a somewhat enhanced $\bar B\to D^{(*)} \mu \bar \nu_\mu$ must be there if one tries to solve the $b\to s \mu^+\mu^-$ anomalies simultaneously. Although such a large deviation has not been established, theoretical calculations for 
the absolute values for the SM predictions and the experimental measurements may have some errors, so a certain level of deviation can be tolerated. We will take a conservative attitude to only allow up to 10\% deviation from SM value, 
in $R^{SM}_\mu(c)$. We find that even such modest requirement put stringent constraint and making the attempt of simultaneously solve the two types of anomalies difficult.

Defining $r(\bar B \to D^{(*)} \tau \bar \nu) = R(\bar B \to D^{(*)} \tau \bar \nu)/R(\bar B \to D^{(*)} \tau \bar \nu)_{SM}$, we have
\begin{eqnarray}
&&r(\bar B \to D^{(*)} \tau \bar \nu) = {2R^{SM}_\tau(c)\over R^{SM}_\mu(c)+R^{SM}_e(c)}\;.
\label{BD}
\end{eqnarray}

Changing $c$ to $u$, one can obtain the R-parity vilating contributions to $R(\bar B \to (\rho, \pi) \tau \nu)$. With the same approximation as above, we have
\begin{eqnarray}
&&r(\bar B \to \tau \bar \nu)=r(\bar B \to (\rho, \pi) \tau \bar \nu)={2R^{SM}_\tau(u)\over R^{SM}_\mu(u)+R^{SM}_e(u)}\;.\label{Bp}
\end{eqnarray}

The linear terms in $r(\bar B \to D^{(*)} \tau \bar \nu)$ and  $r(\bar B \to \tau \bar \nu)$ are proportional to
\begin{eqnarray}
(2 \lambda^{\prime}_{33k}\lambda^{\prime *}_{31k} - \lambda^{\prime}_{23k}\lambda^{\prime *}_{21k}){V_{cd}\over V_{cb}} + (2\lambda^{\prime}_{33k}\lambda^{\prime *}_{32k}- \lambda^{\prime}_{23k}\lambda^{\prime *}_{22k}){V_{cs}\over V_{cb}} +(2 \lambda^{\prime}_{33k}\lambda^{\prime *}_{33k} - \lambda^{\prime}_{23k}\lambda^{\prime *}_{23k})\nonumber
\end{eqnarray}
and 
\begin{eqnarray}
(2 \lambda^{\prime}_{33k}\lambda^{\prime *}_{31k} - \lambda^{\prime}_{23k}\lambda^{\prime *}_{21k}){V_{ud}\over V_{ub}} + (2\lambda^{\prime}_{33k}\lambda^{\prime *}_{32k}- \lambda^{\prime}_{23k}\lambda^{\prime *}_{22k}){V_{us}\over V_{ub}} +(2 \lambda^{\prime}_{33k}\lambda^{\prime *}_{33k} - \lambda^{\prime}_{23k}\lambda^{\prime *}_{23k})\;,\nonumber
\end{eqnarray}
respectively. Note that there is a large enhancement factor $(V_{ud}/V_{ub})/(V_{cb}/V_{cd})$ for the first term in the expression for $r(\bar B \to \tau \bar \nu)$ compared with $r(\bar B \to D^{(*)}\tau \bar \nu)$. This may cause potential problem for a small deviation from 1 in $r(\bar B \to D^{(*)}\tau \bar \nu)$ to a large deviation in $r(\bar B \to \tau \bar \nu)$. One can avoid such a large enhancement by  setting $\lambda^{\prime}_{31k, 21k}$ to be much smaller than other terms.  In our later discussions we will set $\lambda^{\prime}_{31k}$ to be zero. The $\lambda^{\prime}_{21k}$ is also constrained to be small from $D^0 \to \mu^+\mu^-$ decay to be discussed in the following. But may play some important role in $b\to s \mu^+\mu^-$ decay. We will keep it in our discussions.

The SM predictions and experimental measurements for $R(D^{(*)})$ are\cite{HFAG}
\begin{eqnarray}
&&R(D)_{SM} = 0.300\pm 0.008\;, \;\;\;\;\;R(D) = 0.397 \pm 0.040 \pm 0.028\;,\nonumber\\
&&R(D^*)_{SM} = 0.252\pm 0.003\;, \;\;\;\;R(D^*) = 0.316 \pm 0.016 \pm 0.010\;.
\end{eqnarray}

The R-parity violating contributions to both $R(D)$ and $R(D^*)$ occur in a similar way, we use the averaged $r(\bar B \to D^{(*)} \tau \bar \nu)_{ave} = 1.266\pm 0.070$ of $r(\bar B\to D \tau \bar \nu)$ and $r(\bar B \to D^* \tau \bar \nu)$ to represent the anomaly. In the SM, $r_{ave} = 1$. To obtain a $r_{ave}$ within the $1\sigma$ region, $\lambda^\prime_{33k}$ is typically of order  $\sim 3$. This large coupling makes it worrisome for this scenario from unitarity consideration.  In more general terms, the unitarity
limits concern the upper bound constraints on the coupling constants imposed by the condition
of a scale evolution between the electroweak and the unification scales, free of divergences
or Landau poles for the entire set of coupling constants. If so, the R-parity couplings are constrained to be about one at TeV scale\cite{R-parity}. A value of 3 is not consistent. The requirement of no Landau pole up to unifications scale may be not necessary if some new physics appear. One cannot for sure rule out the possibility of reaching unitarity bound of $\sqrt{4\pi}$ at a lower energy. However when attempt to also solve
$b\to s \mu^+\mu^-$ induced anomalies, the model become much more constrained.

\section{Constraints from other tree level processes}

Several other rare processes may receive tree level R-parity violating contributions. The constraints from these processes should be taken into account.
We now study a few of the relevant ones: $K \to \pi \nu\bar \nu$, $\bar B \to K (K^{*}) \nu \bar \nu$, and $D^0 \to  \mu^+\mu^-$. 

The possible terms generating these decays are 
\begin{eqnarray}
&&{\lambda'_{ijk}\lambda^{'*}_{i'j'k}\over 2 m^2_{\tilde
d^k_R}} \bar \nu^{i'}_L \gamma^\mu \nu^i_L \bar d^{j'}_L
\gamma_\mu d_L^j \;,\;\;{\lambda'_{ijk}\lambda^{'*}_{i'j'k}\over 2 m^2_{\tilde
d^k_R}}  \bar e^{i'}_L \gamma^\mu e^i_L 
(\bar u_L V^{KM})^{j'}\gamma_\mu (V^{KM\dagger}u_L)^j \;,\nonumber\\
&&{\lambda'_{ijk}\lambda^{'*}_{i'jk'}\over 2m^2_{\tilde d^j_L} }
\bar \nu^{i'}_L \gamma^\mu \nu^i_L \bar d^k_R \gamma_\mu d^{k'}_R\;,
\;\;{\lambda'_{ijk}\lambda^{'*}_{i'jk'}\over 2m^2_{\tilde u^j_L}}
\bar e^{i'}_L \gamma^\mu e^i_L \bar d^k_R \gamma_\mu
d^{k'}_R\;.
\end{eqnarray}

If $\lambda^\prime_{ijk}$ is non-zero for $k$ restricted to only one value, the two terms on the second line in the above equation will not induce the decays in question. For simplicity, we will work with this assumption \footnote{   
If $k$ can take more than one values, to avoid potential problems from other terms in Eq.(10), one may resort to the scenario
that $\tilde d_L$, $\tilde u_L$, $\tilde e_L$, and $\tilde \nu_L$ to be much heavier than $\tilde d_R$ so that their contributions are suppressed. }.

$D^0\to \mu^+\mu^-$ decay in the SM is extremely small. In our case, there are tree contributions which are therefore constrained severly. We have
\begin{eqnarray}
&&H_{eff} = -  {1\over 2 m^2_{\tilde d^k_R}} C^k_{D\mu\mu}\mu_L\gamma_\mu \mu_L \bar u_L \gamma^\mu c_L\;,\nonumber\\
&&C^k_{D\mu\mu} = \lambda^\prime_{2jk}\lambda^{\prime *}_{2j'k}V_{1j'}V^*_{2j}\nonumber\\
&&= ( \lambda^\prime_{21k}V^*_{21} + \lambda^\prime_{22k}V^*_{22} +  \lambda^\prime_{23k}V^*_{23} )(\lambda^{\prime *}_{21k}V_{11} + \lambda^{\prime *}_{22k}V_{12} + \lambda^{\prime *}_{23k}V_{13})\;.
\end{eqnarray}

The decay width is given by
\begin{eqnarray}
\Gamma(D^0\to \mu^+\mu^-) = {1 \over 128 \pi }  \left \vert {C^k_{D\mu\mu}\over m^2_{\tilde d^3_R}} \right \vert ^2
f_D^2 m_D m^2_\mu  \sqrt{1-{4 m^2_\mu\over m^2_D}}\;,
\end{eqnarray}
where $f_D = 212(1)$ MeV\cite{fD} is the $D^0$ decay constant.

Using experimental upper bound\cite{PDG} $6.2\times 10^{-9}$ at 90\% C.L. for $D^0\to \mu^+\mu^-$, we have
$\vert C^k_{D\mu\mu}{(1\mbox{TeV})^2 /m^2_{\tilde d^k_R}}\vert < 6.1\times 10^{-2}$.
With $\lambda^\prime_{21k, 22k}$ set to zero, $C^k_{D\mu\nu}$ is give by
$C^k_{D\mu\mu} = \lambda^{\prime}_{23k}\lambda^{\prime *}_{23k}V_{ub}V_{cb}^*$. 
We have $\lambda^\prime_{23k} \lambda^{\prime *}_{23k}(1\mbox{TeV})^2 / m^2_{\tilde d^k_R} < (20)^2$.
$\lambda^\prime_{23k}$ is only very loosely constrained from $D^0\to \mu+\mu^-$. If just $\lambda^\prime_{21k}$ or $\lambda^\prime_{22k}$ is non-zero, they are constrained as
\begin{eqnarray}
 \lambda^\prime_{21k} \lambda^{\prime *}_{21k}{(1\mbox{TeV})^2 \over m^2_{\tilde d^k_R}},  \lambda^\prime_{22k} \lambda^{\prime *}_{22k}{(1\mbox{TeV})^2 \over m^2_{\tilde d^k_R}} < 0.28\;.\label{dmumu}
 \end{eqnarray}
These constraints on  $\lambda^\prime_{21k}$ and $\lambda^\prime_{22k}$, make their effects on $b\to s\mu^+\mu^-$ small. 
Later we will show that even a small $\lambda^\prime_{22k}$ may play some important role in having a better coherent explanation of $R(D^{(*)})$ and $b\to s \mu^+\mu^-$ anomalies.

For $K \to \pi \nu\bar \nu$, the ratio of $R_{K\to \pi \nu\bar \nu} =\Gamma_{RPV}/\Gamma_{SM}$ is given by\cite{ddh}
\begin{eqnarray}
&&R_{K\to \pi \nu\bar \nu}  = \sum_{i=,e,\mu,\tau}{1\over 3}\left \vert 1 +{\Delta^{RPV}_{\nu_i\bar \nu_i}\over X_0(x_t) V_{ts}V^*_{td}} \right \vert ^2 + {1\over 3} \sum_{i\neq i'} \left \vert {\Delta^{RPV}_{\nu_i\bar \nu_{i'} }\over X_0(x_t)V_{ts}V^*_{td} }\right \vert ^2\;,\nonumber\\
&&\Delta^{RPV}_{\nu_i \bar \nu_{i'}} = {\pi s^2_W \over \sqrt{2} G_F \alpha} \left \vert -{\lambda^{\prime}_{i2k}\lambda^{\prime *}_{i' 1 k}
\over 2 m^2_{\tilde d^k_R}} \right \vert ^2\;,\;\;\;\;X_0(x) = {x(2+x)\over 8(x - 1)} + {3x(x-2)\over 8(x-1)^2}\ln x\;,
\end{eqnarray}
where $x_t = m^2_t/m^2_W$. 

Combining the SM prediction\cite{kpiuu-sm} for the branching ratio  and experimental information\cite{PDG}
$Br = (1.7\pm1.1)\times 10^{-10}$, at $2\sigma$ level, 
$\lambda^{\prime}_{i2k}\lambda^{\prime *}_{i'1k}$ are constraint to be less than a few times of $10^{-3} (m^2_{d^k_R}/ (1\mbox{TeV})^2)$.
Since we will set $\lambda^{\prime *}_{i1k}=0$, this process is not affected at tree level. 

The expressions  for $R_{\bar B \to \pi \nu\bar \nu}$ and $R_{\bar B \to K(K^*)\nu\bar \nu}$ of $\bar B\to \pi \nu\bar \nu$ and $\bar B \to K(K^*)\nu\bar \nu$ can be obtained from Eq.(14) by replacing $V_{ts}V^*_{td}$ to $V_{tb}V^*_{td}$ and $V_{tb}V^*_{ts}$, respectively. The corresponding $\Delta^{RPV}_{\nu_i \bar \nu_{i'}}$ are
\begin{eqnarray}
&&\mbox{For}\; \bar B\to \pi \nu \bar \nu: \;\;\;\;\;\;\;\;\;\;\Delta^{RPV}_{\nu_i \bar \nu_{i'}} = {\pi s^2_W \over \sqrt{2} G_F \alpha} \left \vert -{\lambda^{\prime}_{i3k}\lambda^{\prime *}_{i' 1 k}
\over 2 m^2_{\tilde d^k_R}} \right \vert ^2\;,\nonumber\\
&&\mbox{For}\; \bar B\to K(K^*) \nu \bar \nu: \;\;\Delta^{RPV}_{\nu_i \bar \nu_{i'}} = {\pi s^2_W \over \sqrt{2} G_F \alpha} \left \vert -{\lambda^{\prime}_{i3k}\lambda^{\prime *}_{i' 2 k}
\over 2 m^2_{\tilde d^k_R}} \right \vert ^2\;.
\end{eqnarray}

For $B\to \pi \nu\bar \nu$, since we have set $\lambda^\prime_{i1k} = 0$, it is again not affected by R-pairty violating interactions in this model. 

The process $\bar B\to K(K^{*}) \nu\bar \nu$ will be affected. We have the following non-zero $\Delta^{RPV}_{\nu\bar\nu}$
\begin{eqnarray}
&&\Delta^{RPV}_{\nu_\mu\bar \nu_\mu} = -{\lambda^{\prime}_{23k}\lambda^{\prime *}_{22k} \over 2 m^2_{d^k_R}}{\pi s^2_W\over \sqrt{2} G_F \alpha}\;,\;\;\Delta^{RPV}_{\nu_\tau\bar \nu_\tau} = -{\lambda^{\prime}_{33k}\lambda^{\prime *}_{32k} \over 2 m^2_{d^k_R}}{\pi s^2_W\over \sqrt{2} G_F \alpha}\;,\nonumber\\
&&\Delta^{RPV}_{\nu_\tau\bar \nu_\mu} = -{\lambda^{\prime}_{33k}\lambda^{\prime *}_{22k} \over 2 m^2_{d^k_R}}{\pi s^2_W\over \sqrt{2} G_F \alpha}\;,\;\;
\Delta^{RPV}_{\nu_\mu\bar \nu_\tau} = -{\lambda^{\prime}_{23k}\lambda^{\prime *}_{32k} \over 2 m^2_{d^k_R}}{\pi s^2_W\over \sqrt{2} G_F \alpha}\;. \label{bnunu}
\end{eqnarray}

Experimental data from BaBar\cite{Babar} and Belle\cite{Belle} give, $R_{B\to K(K^*) \nu\bar \nu} <4.3 (4.4)$ implying
$\lambda^{\prime}_{23k}\lambda^{\prime *}_{22k}$, $\lambda^{\prime}_{33k}\lambda^{\prime *}_{32k}$,  $\lambda^{\prime}_{33k}\lambda^{\prime *}_{22k}$ and $\lambda^{\prime}_{23k}\lambda^{\prime *}_{32k}$
are constrained from $\bar B\to K(K^*)\nu\bar \nu$. We shall return to this process later.

\section{Loop contributions for $b\to s \mu^+\mu^-$ induced anomalies}

The anomalous effects in $b\to s\mu^+\mu^-$ induced processes are only 2 to 3 $\sigma$ effects and need to be confirmed further. They may be due to our poor understanding of  hadronic matrix elements involved, and may also be caused by new physics beyond SM. We now discuss how R-parity violating interaction may help to solve the problems.

New physics contributes to $b\to s l \bar l$ can be parametrized as  $H^{NP}_{eff} = \sum C^{NP}_i O_i$. Some of the most studied operators $O_i$ are
\begin{eqnarray}
&&O_9 = {\alpha\over {4\pi} }\bar s \gamma^\mu P_L b \bar \mu\gamma_\mu \mu\;,\;\;\;\;\;\;\;\;\;O^{\prime}_9 = {\alpha\over {4\pi}} \bar s \gamma^\mu P_R b \bar \mu\gamma_\mu \mu\;,\nonumber\\
&&O_{10} = {\alpha \over {4\pi}} \bar s \gamma^\mu P_L b \bar \mu\gamma_\mu \gamma_5 \mu\;,\;\;\;\;O^{\prime}_{10} = {\alpha \over {4\pi}} \bar s \gamma^\mu P_R b \bar \mu\gamma_\mu \gamma_5 \mu\;,
\end{eqnarray}
where $P_{L,R} = (1\mp \gamma_5)/2$.

The SM predictions are $C^{SM}_9 \approx - C^{SM}_{10} = 4.1$. A global analysis shows that to solve the anomalies in decays induced by $b\to s \mu^+\mu^-$, there are few scenarios where the anomalies can be solved with high confidence level and all cases $C^{NP}_9$ need to be around $-1$\cite{b2s-sm}. For example with $C^{NP}_9 = -1.09$ and $C^{NP}_{10}$, $C^{\prime, NP}_{9,10} = 0$ with a 4.5 pull; the cases with  $C^{NP}_9 = - C^{\prime, NP}_{9}$, the best fit values are:
$C^{NP}_{9} = -C^{\prime, NP}_{9} = -1.06$ and others equal to zero with a 4.8 pull; And the case with $C^{NP}_9 = - C^{NP}_{10}$, the best fit values are:
$C^{NP}_{9} = -C^{NP}_{10} = -0.68$ and others equal to zero with a 4.2 pull. Here the number of  ``pulls'' indicates by how many sigmas the best fit point is preferred over the SM point for a given scenario. The higher the pull,  the better fit between theory and experimental data is reached. In our case, the R-parity violating contribution to be discussed belongs to the last case. For this case, the $1\sigma$ allowed range is\cite{b2s-sm},  $-0.85 \sim -0.5$. With negative value for $C^{NP}_9$, the new physics contribution reduces $b\to s \mu^+\mu^-$ and therefore helps to explain why $B \to (K,K^*,\phi)\mu^+\mu^-$ branching ratios and $R_K$ are smaller than those predicted by  SM.

There is a potential contribution to $b\to s \mu^+\mu^-$ at tree level due to a term proportional to 
$\lambda'_{ijk}\lambda^{'*}_{i'jk'}/2m^2_{\tilde u^j_L} \bar e^{i'}_L \gamma^\mu e^i_L \bar d^k_R \gamma_\mu d^{k'}_R$.
However, since we assume that there is only one non vanishing value for $k$, $b\to s \mu^+\mu^-$ is not induced by this contribution. 

One needs to include one loop contributions.
At one loop level, exchanging $\tilde d^k_R$ in the loop, contributions with $C^{NP}_{9} = -C^{NP}_{10}$ can be generated with\begin{eqnarray}
C^{NP,l \bar l'}_{9} &\approx& {m^2_q\over 8\pi \alpha} {1\over m^2_{\tilde d^k_R}}
\lambda^{\prime}_{ l b k}\lambda^{\prime *}_{\bar l' m k} {V_{qm}V^{ *}_{ts}\over V_{tb} V^{ *}_{ts}}\nonumber\\
&-&{\sqrt{2}\over 64 \pi \alpha G_F} 
{\ln (m^2_{\tilde d^k_R}/m^2_{\tilde d^{k'}_R}) \over m^2_{\tilde d^k_R} - m^2_{\tilde d^{k'}_R}}
\lambda^{\prime}_{ibk}\lambda^{\prime *}_{is k'} \lambda^{\prime}_{l j k'}\lambda^{\prime *}_{\bar l' j k}
{1 \over V_{tb} V^{*}_{ts}}\;,
\end{eqnarray}
where $m_q$ is the up type quark mass. The first term is induced by exchanging a $W$ boson and a sparticle $\tilde d^k_R$, and the second term is by exchanging two sparticles $\tilde d^k_R$ in the loops. The term of interest corresponds to $l=2$, $\bar l'=2$, $s=2$ and $b=3$ for the process $b\to s \mu^+ \mu^-$. One can relabel them with different numbers for other process.

The first term is dominated by $q=t$, its contribution to $C_9^{NP,\mu\bar \mu}$ is about $0.15 \lambda^\prime_{23k}\lambda^{\prime *}_{23k}(1\mbox{TeV}/m_{\tilde d^k_R})^2$. This is a ``wrong sign'' contribution to solve  $b\to s\mu^+\mu^-$ induced anomalies\footnote{In our earlier version, we had neglected this contribution and obtained erroneous  conclusions which we correct them here. }. With $\lambda^\prime_{1jk} = 0$ and $\lambda^\prime_{i1k} =0$ from considerations of no processes with electron has shown anomalies and $K\to \pi \nu \bar \nu$ constraint, and restricting $k$ to have only one value, we have 
\begin{eqnarray}
C^{NP, l \bar l'}_{9} &\approx& {m^2_t\over 8\pi\alpha} {1\over m^2_{\tilde d^k_R}}
\lambda^{\prime}_{ l3 k}\lambda^{\prime *}_{\bar l' 3 k} \nonumber \\
&-&{\sqrt{2}\over 64 \pi \alpha G_F} 
{1\over m^2_{\tilde d^3_R}}
(\lambda^{\prime}_{23k}\lambda^{\prime *}_{22 k}+ \lambda^{\prime}_{33k}\lambda^{\prime *}_{32 k})(\lambda^\prime_{l2k}\lambda^{\prime *}_{\bar l' 2k}+\lambda^\prime_{l3k}\lambda^{\prime *}_{\bar l' 3k} )
{1 \over V_{tb} V^{*}_{ts}}\\
&=& \left (0.157 
\lambda^{\prime}_{ l3 k}\lambda^{\prime *}_{\bar l'3 k} +2.0
(\lambda^{\prime}_{23k}\lambda^{\prime *}_{2 2 k}+ \lambda^{\prime}_{33k}\lambda^{\prime *}_{32 k})(\lambda^\prime_{l2k}\lambda^{\prime *}_{\bar l'2k}+\lambda^\prime_{l 3k}\lambda^{\prime *}_{\bar l' 3k} )\right ){(1\mbox{TeV})^2\over m^2_{\tilde d^k_R}}\;.\nonumber \label{c9t}
\end{eqnarray}

\section{Numerical analysis}

We are now in a position to put things together to see if R-parity violating interactions may be able to solve the $R(D^{(*)})$ and $b\to s \mu^+\mu^-$ anomalies simultaneously. For the KM parameters we use those given in Particle Data Group\cite{PDG}. The aim is to produce values for $r(\bar B\to D^{(*)}\tau \bar \nu)_{ave}$, $C_{9}^{NP}$ as close as possible to their central values, $1.266$ and $-0.68$.  At the same time we have to restrict $R_{\bar B \to K(K^*)\nu \bar \nu}$ to be less than 4.3 to satisfy experimental bound. 

If one just needs to solve the $R(D^{(*)})$ anomaly, one just can easily obtain the central value of $r_{ave}-1 = 0.266$ by setting all other $\lambda^\prime_{ijk}$ to zero except $\lambda^\prime_{33k}$ with its vale given by $2.95(m_{\tilde d^k_B}/1\mbox{TeV})$. If $m_{\tilde d^k_R}$ is way above TeV, then the coupling will violate the unitarity bound of $\sqrt{4\pi}$. Therefore for the theory to work purterbatively , one expect the squark mass to be less than a TeV or so which can be looked for at the LHC. With this choice of $\lambda^\prime$ the SM predictions for $R_{\bar B\to K \nu\bar \nu,\;K\to \pi \nu \bar \nu}$ and $\Gamma(D^0\to \mu^+\mu^-)$ will not be affected, and $R^{SM}_{e,\mu}(c,u) = 1$. One also predicts $r(\bar B \to D^{(*)}\tau \bar \nu) = r(\bar B\to \tau \bar \nu)=1.26$. This can be tested by future experimental data. This is the scenario discussed in Ref.\cite{desh-menon}. One can try to ease the unitarity bound  by including the 
$\lambda^\prime_{33k}\lambda^{\prime *}_{32k}$ term with positive sign so that a smaller $\lambda^\prime_{33k}$ value is now allowed.

 We now discuss the contributions to $C^{NP}_9$ from eq.(\ref{c9t}). Note that the first term in that equation is positive definate, one needs a larger second term with negative sign to produced the required value. If one just needs to satisfy this equation, one can easily find solutions. 
For example, taking $\lambda^\prime_{23k} = 3.0$, one just needs to have $\lambda^\prime_{23k}\lambda^{\prime*}_{22k} + \lambda^\prime_{33k}\lambda^{\prime*}_{32k}$ to be about -0.046 to produce $C^{NP}_9\sim -0.68$. 

One, however, has to consider other strong constraints. A particularly important constraint is from eq.(\ref{bnunu}), to satisfy $R_{\bar B\to K \nu\bar \nu} <4.3$. To produce a negative $C^{NP}_9$,  $\lambda^\prime_{23k}\lambda^{\prime*}_{22k} + \lambda^\prime_{33k}\lambda^{\prime*}_{32k}$ needs to be negative. From $R_{\bar B \to K \nu \bar \nu}$ constraint, each of $r_{2322}=\lambda^\prime_{23k}\lambda^{\prime*}_{22k}$ and $r_{3332}=\lambda^\prime_{33k}\lambda^{\prime*}_{32k}$ is constrained by be larger than $-0.09$. But in general they appear together in order to produce the value required for $C^{NP}_9$. This also leads to non-zero values for $r_{2332}=\lambda^\prime_{23k}\lambda^{\prime*}_{32k}$ and $r_{3322}=\lambda^\prime_{33k}\lambda^{\prime*}_{22k}$ increasing the value for $R_{\bar B \to K \nu \bar \nu}$. We find that with all $r_{ijkl}=-0.0436$ having the same value maximizes the size of $C^{NP}_9$ while minimize $R_{\bar B\to K\nu \bar \nu}$. For this case, using $C^{NP}_9 = -0.68$ and $R_{\bar B \to K \nu \bar \nu} <4.3$, we find $\lambda^\prime_{33k,23k} = 6.3$ and a small  value for $\lambda^\prime_{22,32} =-0.0068$. With the above values for $\lambda^\prime$, the constraints from $D^0\to \mu^+\mu^-$ can be satisfied. However, the predicted values for $r_{ave}$ becomes 1.48 and $R^{SM}_\mu(c)$ is about 2.9. These values are completely ruled out by existing data. Also the solution with $\lambda^\prime_{33k,23k} = 6.3$ is problematic because it violates the unitarity bound and therefore is not a viable solution neither.  In figure 1, we show  $C^{NP}_9$, $r_{ave}$ and $R^{SM}_{\mu}(c)$ as functions of $\lambda^\prime_{23k}$. We see a smaller $C^{NP}_9$ in size may relax the situation, but within 1$\sigma$ range for $C^{NP}_9$, 
The value for $R^{SM}_{\mu,\tau}(c)$ are too large to allow the model to be a viable one.

\begin{figure}[h]

\includegraphics[scale=0.5]{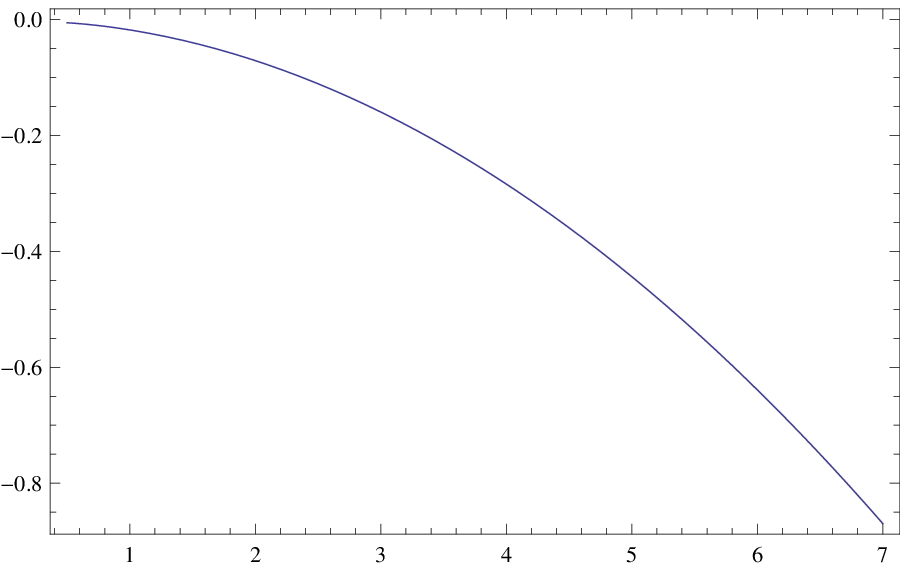} \hspace{0.2cm}
\includegraphics[scale=0.5]{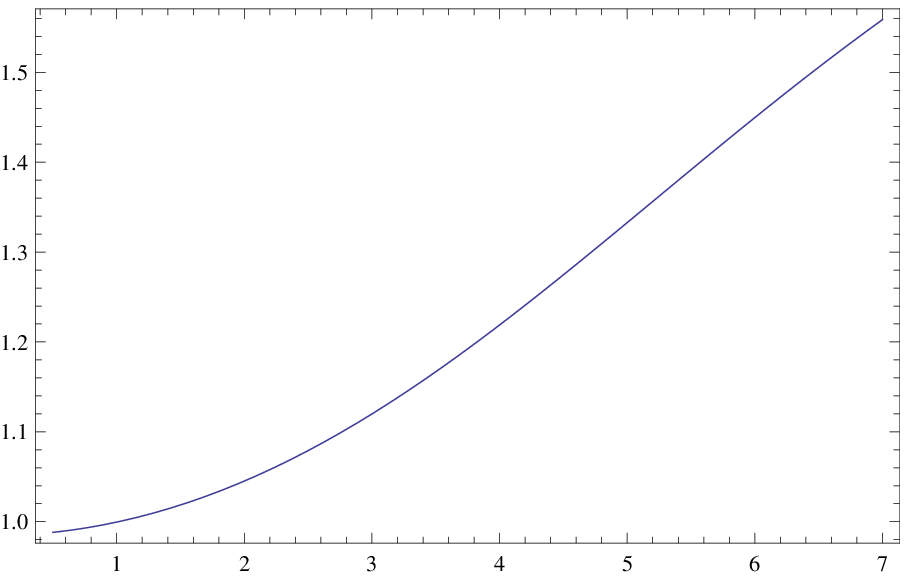} \hspace{0.2cm}
\includegraphics[scale=0.5]{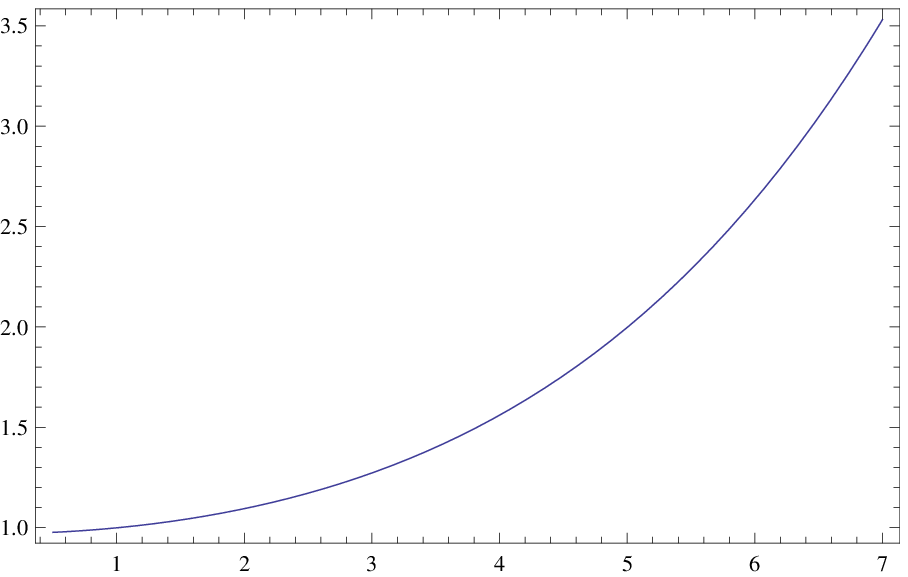} 
\caption{$C^{NP}_9$, $r_{ave}$ and $R^{SM}_{\mu}(c)$ as functions of $\lambda^\prime_{23k}$ from left to right, respectively. 
To get $R^{SM}_{\mu}(c)-1$ down to 10\%, one needs to go to the lower range the $3\sigma$ range for $C^{NP}_9$ to about -0.18\cite{b2s-sm}. However, in that case, $r_{ave}$ also comes down and cannot explain the observed $R(D^{(*)})$ anomaly.}
\label{fig}
\end{figure}

We have searched a wide range of parameter space for $\lambda^\prime$ including with complex numbers and found no solutions which can simultaneously satisfy 
bounds on $R_{\bar B \to K\nu\bar \nu}$ and $R^{SM}_\mu(c)$ and 
at the same time to solve anomalies in $R(D^{(*)})$ and $b\to s \mu^+\mu^-$.

\section{Conclusions}

We have studied the possibility of explaining the enhancement in semileptonic decays of $\bar B \to D^{(*)} \tau \bar \nu$ and the anomalies induced by $b\to s\mu^+\mu^-$ within the framework of R-parity violating (RPV) MSSM. Exchange of down type right-handed squark coupled to quarks and leptons yield  interactions which are similar to leptoquark induced interactions which have been proposed to explain the $\bar B \to D^{(*)}\to \tau \bar \nu$ by tree level interactions and $b\to s \mu^+\mu^-$ induced anomalies by loop interactions, simultaneously. However, we find that the Yukawa couplings have severe constraints from other rare processes in $B$ and $D$ decays. This interaction can provide a viable solution to $R^{D(*)}$ anomaly. But with the severe constraint from 
$\bar B \to K \nu \bar \nu$, it proves impossible to solve the anomalies induced by $b\to s \mu^+\mu^-$. This conclusion also applies equally to the leptoquark model proposed in Ref.\cite{bauer-neubert}.

\begin{acknowledgments}

This work was supported by a University of Oregon Global Studies Institute grant awarded to NGD and XGH. XGH was supported in part by MOE Academic Excellent Program (Grant No.~102R891505), NCTS and MOST of ROC (Grant No.~MOST104-2112-M-002-015-MY3), and in part by NSFC (Grant Nos.~11175115 and 11575111) and Shanghai Science and Technology Commission (Grant No.~11DZ2260700) of PRC.  XGH thanks the Institute of Theoretical Science, Department of Physics, University of Oregon for hospitality where this work was done. We thank M. Schmidt for bringing Ref.\cite{new} to our attention.
\\
\\
{\bf Note Added}

Please note that soon after our submission Becirevic et al.\cite{new}
have submitted a paper to arXiv which reaches similar conclusion on the
inadmissibility of single leptoquark explanation of anomalies.

\end{acknowledgments}

\end{document}